\documentclass[twocolumn,prd,showpacs,10pt]{revtex4}
\usepackage{graphicx}
\usepackage{amsmath}
\usepackage{amssymb}
\usepackage{amsfonts}
\usepackage{lscape}
\usepackage{epsfig}

\usepackage{color}
\usepackage{bm}
\def\be{\begin{equation}}
\def\ee{\end{equation}}
\def\bea{\begin{eqnarray}}
\def\eea{\end{eqnarray}}
\newcommand{\comment}[1]{}
\def\Om{\Omega_{m_0}}
\def\css{c_s^2}
\def\cv{c_{vis}^2}

\newcommand{\lsim}{\,\raise 0.4ex\hbox{$<$}\kern -0.8em\lower 0.62ex\hbox{$\sim$}\,}
\newcommand{\gsim}{\,\raise 0.4ex\hbox{$>$}\kern -0.7em\lower 0.62ex\hbox{$\sim$}\,}
\newcommand{\ca}{{c_a^2}}
\newcommand{\cs}{{c_s^2}}

\newcommand{\dep}{\delta p}

\newcommand{\lakr}{\left(a,k\right)}
\newcommand{\lkr}{\left(k\right)}
\newcommand{\lar}{\left(a\right)}
\def\AA{{\cal A}}
\def\ceff{{c_{eff}^2}}

\begin{document}

\title{Fingerprinting Dark Energy III: distinctive marks of viscosity}

\author{Domenico Sapone}
\email{domenico.sapone@uam.es}
\affiliation{Instituto de F\'\i sica Te\'orica (UAM/CSIC), Universidad Aut\'onoma de Madrid, Cantoblanco 28049 Madrid, Spain}
\author{Elisabetta Majerotto}
\email{elisabetta.majerotto@uam.es}
\affiliation{Instituto de F\'\i sica Te\'orica (UAM/CSIC) \& Departamento de F\'\i sica Te\'orica (UAM), 
Universidad Aut\'onoma de Madrid, Cantoblanco 28049 Madrid, Spain}

 \begin{abstract}
 The characterisation of dark energy is one of the primary goals in cosmology especially now that many new 
 experiments are being planned with the aim of reaching a high sensitivity on cosmological parameters. 
 It is known that if we move away from the simple cosmological constant model then we need to consider perturbations 
 in the dark energy fluid. This means that dark energy has two extra degrees of freedom: 
 the sound speed $\cs$ and the anisotropic stress $\sigma$.  
 If dark energy is inhomogenous at the scales of interest then the gravitational potentials are modified and 
 the evolution of the dark matter perturbations is also directly affected. 
 In this paper  we add an anisotropic component to the 
 dark energy perturbations. Following the idea introduced in \cite{Sapone:2009mb}, 
 we solve analytically the equations of perturbations in the dark sector, finding simple and accurate approximated solutions. 
 We also find that the evolution of the density perturbations is governed by an {\em effective sound speed} which depends on 
 both the sound speed and the anisotropic stress parameter. 
 We then use these solutions to look at the impact of the dark energy perturbations on the matter power spectrum and on the 
 Integrated Sachs-Wolfe effect in the Cosmic Microwave  Background. 
 \end{abstract}

\date{\today}

\maketitle

\section{Introduction}

The cause of the acceleration of the Universe's expansion, real or apparent it may be,  is yet shrouded in mystery. 
The easiest explanation involves Einstein's cosmological constant and results in the ``standard" cosmological 
model, $\Lambda$CDM. However, this suffers of fine tuning problems, 
as well as all up to now available alternative models.
Since the observations of supernovae type Ia (SNIa)  \cite{sn1,sn2}, many other probes have confirmed 
this acceleration, and several experiments are planned in order to understand the nature of the 
phenomenon, usually called ``dark energy''. The quality and quantity of the upcoming data will allow 
to better distinguish among different models, both from the background expansion and from the 
cosmological perturbations point of view. In this context, it is important to determine and understand 
as well as possible all signatures characterising different models.

In two previous papers \cite{Sapone:2009mb, Sapone:2010uy}, some of us have studied 
the influence of dark energy perturbations on the evolution of cosmological perturbations 
and on the observables related to it. To quantify this impact, the two parameters 
$Q$ and $\Sigma$ introduced in \cite{Amendola:2007rr} were evaluated, the first measuring 
the fraction of dark energy  density perturbations and the second related to the sum of 
the gravitational potentials, observable with weak lensing. In both \cite{Sapone:2009mb, Sapone:2010uy}, 
dark energy was modeled as a fluid characterised by an equation of state 
$w = p/\rho$ (where $p$ and $\rho$ are pressure and energy density, respectively) 
and a sound speed $\cs = \delta p/\delta\rho$ (where $\delta p$ are the pressure perturbations), 
both assumed to be constant. Paper \cite{Sapone:2009mb} found a simple and useful 
analytical expression for matter and dark energy perturbations, 
while paper \cite{Sapone:2010uy} looked more carefully at observational implications.

Here we take a further step in generalising the dark energy fluid. We introduce an additional 
degree of freedom: the anisotropic stress. From the seminal work \cite{Kodama:1985bj}, 
we know that a classical scalar field has no anisotropic stress. Although most dark energy 
models are based on scalar fields (for a review see e.g.  \cite{Copeland:2006wr, Sapone:2010iz}), 
it is nevertheless interesting, also based of the fact that very little is known about dark energy, 
to study this very general fluid and see how such a term would affect density perturbations 
and observables which are sensitive to them. Moreover, modified gravity models may be 
reformulated as {\em effective dark energy fluids} with anisotropic stress. The possibility of 
detecting it is therefore related to the problem of distinguishing dark energy from modified gravity.

Probably due to the absence of anisotropic stress in classical scalar fields, which are among the most 
popular candidates for dark energy, there is not much literature on this subject. One interesting paper 
on this is \cite{Hu:1998kj}. Here a ``generalised dark matter" with an anisotropic stress 
component was studied. This anisotropy corresponds, in the case of the fluid, to a viscosity 
term $\cv$ damping density perturbations. The authors of this paper designed an equation 
governing the evolution of the anisotropic stress which recovers the free streaming equations 
of motion for radiation up to the quadrupole. In \cite{Koivisto:2005mm} this same general 
parameterisation of the dark energy component was analysed with data from the cosmic 
microwave background radiation (CMB), large scale structure and supernovae type Ia, 
finding that both $\cs$ and $\cv$ were hard to constrain and that even future data would 
not improve very much in measuring them. A similar conclusion was reached by  \cite{Calabrese:2010uf}. 
Here forecasts were made on how well future CMB experiments will constrain an 
{\em early, cold and stressed} dark energy.  Also \cite{Archidiacono:2011gq} used this 
parameterisation to constrain extra neutrino species which are not explained by particles physics. 
A different approach on anisotropic stress was taken by \cite{Ballesteros:2011cm}, which 
found general consistency relations for $\Lambda$CDM and studied anisotropies arising even in 
$\Lambda$CDM from second order perturbations. A non exhaustive list of other works concerning 
observable consequences of an anisotropic stress term is \cite{Pogosian:2010tj,Silvestri:2009jw}.

In this context, our approach is similar as in \cite{Sapone:2009mb}: we aim to try and solve {\em analytically} 
the perturbation equations for the simplest possible model of an anisotropic stress dark energy fluid in 
the simplest sensible approximations. 
This is useful both to understand more clearly the signatures induced by viscosity and to build simple tools 
for comparisons with observations. 
We parameterise  the dark energy fluid by using three constants: 
the equation of state parameter $w$, the sound speed $\cs$ and the
viscosity $\cv$. We start off from the anisotropic stress model of  \cite{Hu:1998kj}.

The structure of the paper is the following. After discussing
the perturbation equations, defining our variables and describing the model for anisotropic stress in Sec. \ref{sec:def}, 
we derive simple analytical solutions for the dark energy perturbations in presence of anisotropic stress during
the matter domination era in Sec. \ref{sec:pert}. We also verify that our expressions  are a good fit to the numerical 
solutions obtained using CAMB \cite{camb, camb1}. 
In the same section we study the effect of the anisotropic stress of dark energy
on the evolution of perturbations, concluding that the presence of 
a viscosity  term $\cv$ produces an {\em anisotropic} horizon, which 
results in an {\em effective sound speed}. 
In Sec. \ref{sec.QS} we compute the clustering parameters
$Q(k,t)$, $\Sigma(k,t)$ and $\eta(k,t)$ defined in \cite{Amendola:2007rr}, for the case of our model. 
These have the advantage of tracking the numerical solution even after matter domination, where they are strictly valid.
Finally, in Sec. \ref{sec:obs} we use our results to evaluate the effect of the
viscosity $\cv$ (and of the other dark energy parameters $w$ and $\cs$) on the amplitude 
and shape of the matter power spectrum, on the growth factor and on the integrated Sachs-Wolfe (ISW) effect.
We stress that our aim is not to give a full theoretical analysis on imperfect fluid dark 
energy, but to analyse the effect of a possible viscous dark energy on the growth of perturbations.

\section{First order perturbations in dark energy} \label{sec:def}

\subsection{Definitions} 

In this paper we consider only spatially flat universes, and we use the Newtonian or longitudinal gauge
so that our metric reads
\be
ds^{2} = a^{2} \left[ -\left( 1+2\psi \right) d\tau^{2} + \left( 1-2\phi\right) dx_{i}dx^{i} \right] ,
\label{pert_newton_ds}
\ee
where $a$ is the scale factor,  $\tau$ is the conformal time, $\psi$ and $\phi$ are scalar metric 
perturbations (or potentials) and we do not consider vector and tensor perturbations.
 With $H$ we indicate the Hubble parameter, $H = (1/a) (da/dt)$ computed with respect to the 
 physical time $t$ ($dt=a d\tau$) and an overdot indicates a derivative with respect to $\tau$.
Let us remind that while the gauge choice affects the perturbations on scales larger than
the Hubble horizon, $k\lsim aH$, on much smaller scales the observables are independent of it.

For a generic fluid with constant equation of state parameter $w=p/\rho$, the perturbation equations 
are \cite{mabe, rd}
\bea
\delta' &=& 3(1+w) \phi' - \frac{V}{Ha^2} - 3 \frac{1}{a}\left(\frac{\dep}{\rho}-w \delta \right) \label{delta} \\
V' &=& -(1-3w) \frac{V}{a}+ \frac{k^2}{H a^2} \frac{\dep}{\rho}+(1+w) \frac{k^2}{Ha^2} \psi +\\ \nonumber
&-&(1+w)\frac{k^2}{Ha^2}\sigma  \label{v}
\eea
where $\delta = \delta\rho/\rho$ is the density contrast, $V$ is the velocity perturbation, $\delta p$ is 
the pressure perturbation, $\sigma$ is the anisotropic stress of the fluid and with a prime we indicate a derivative with respect to $a$.
Since we are interested in late times, we will assume that the Universe is filled only by two components: a matter fluid with
$w=\delta p = 0$ and a dark energy fluid, parameterised by a
constant $w \sim -1$, by an anisotropic stress function $\sigma$ and by a sound speed $\cs$\footnote{$\cs$ and $\cv$ 
are expressed in units of speed of light $c=1$} related to pressure perturbation through
\be
\delta p = \cs \rho\delta+\frac{3aH\left(\cs-\ca\right)}{k^2}\rho V 
\ee
where $\ca= \dot{p}/\dot{\rho}$ is the adiabatic sound speed of the fluid. 
Here we restrict ourselves to models with $w\gtrsim-1$ to avoid crossing the phantom divide, 
where the above is not a good parameterisation  \cite{Kunz:2006wc}. In our case the adiabatic sound 
speed is $\ca = w$ since our dark energy fluid has constant $w$.

Given the above assumptions the perturbation equations (\ref{delta}) and (\ref{v}) for our dark energy fluid become
\bea
\delta' &=&  - \frac{V}{Ha^2}\left[1+\frac{9a^2 H^2\left(\cs-w\right)}{k^2}\right] + \nonumber \\
&-&\frac{3}{a}\left(\cs-w\right)\delta+3\left(1+w\right)\phi' \label{eq:deltap} \\
V' &=& -(1- 3 \css) \frac{V}{a} + \frac{k^2 \css\delta}{a^2 H} + \frac{(1+w)k^2}{a^2 H}\left[\psi  -  \sigma\right]. \label{eq:vp}
\eea
The first two Einstein equations (in conformal Newtonian gauge) 
 are \cite{mabe}:
\bea
k^2\phi + 3a^3H^2\left( \phi' + \frac{1}{a}\psi\right) &=& 4\pi G a^2 \sum_i\delta T^0{}_{(i)\,0} \,,\label{ein-cona}\\
k^2 \left(\phi' + \frac{1}{a}\psi \right)&=& \frac{4\pi G}{H} \sum_i\rho_i V_i\,,\label{ein-conb}
\eea
where the sum runs over all types of matter that cluster. We can write Eqs. (\ref{ein-cona}) 
and (\ref{ein-conb}) in one single formula to obtain the Poisson equation
\bea 
k^2\phi  &=& -4\pi G a^2\sum_i\rho_i\left( \delta_i+\frac{3aH}{k^2}V_i\right) =\nonumber \\
&=& -4\pi G a^2\sum_i\rho_i\Delta_i \label{eq.Poisson_general}
\eea
where $\Delta_i$ is the (gauge-invariant) comoving density contrast of the i$^{\rm th}$ fluid.
The fourth Einstein equation, which describes the difference 
between the two potentials $\psi$ and $\phi$ as a function of the anisotropic stress, is
\bea  \label{eq.4thEinsteinB}
k^2\left(\phi -\psi \right) &=& 12 \pi G a^2 \, (1+w) \rho\, \sigma  \\
&=& \frac{9}{2}  H_0^2 (1-\Om)a^{-(1+3 w)} (1+w)  \sigma  \nonumber \\
&\equiv& B(a)\,\sigma \,,\label{eq.4thEinstein}
\eea
where $\rho$ is the energy density of dark energy. To be able to solve analytically the above equations, 
we restrict ourselves to the matter dominated epoch, which is also the most interesting regime 
in relation to observations. Here $8 \pi G \sum_{i}\rho_{i}  \simeq 8\pi G \rho_m  = 3H_0^2 \Om a^{-3}$.
 A consequence of this is that $\sum_{i}\rho_{i}\Delta_{i} \simeq \rho_{m}\Delta_{m}$ so that,  at first approximation, 
 only matter perturbations contribute to source the gravitational potential in 
Eq.~(\ref{eq.Poisson_general}). 
Since we are interested precisely in the impact of a non-vanishing anisotropic stress of dark energy on the main observables,
 we should in principle not neglect $\sigma$ in Eq. (\ref{eq.4thEinsteinB}). 
However, we notice that the term $B(a) $ in Eq.~(\ref{eq.4thEinstein}) 
is proportional to $1+w \sim 0$ (with $w\sim -1$) and to $a^{-(1-3w)} \sim a^{-4}$ which decays away very quickly.
Hence, as a first approximation, we neglect the contribution of $\sigma$ in the fourth Einstein equation,
obtaining $ \psi \simeq  \phi$.

\subsection{Anisotropic stress dark energy}

To model a dark energy anisotropy is difficult as this might rise from 
a real internal degree of freedom of the fluid or from a modification of the geometry of spacetime.
Scalar field dark energy for example always has $\sigma = 0$ \cite{Kodama:1985bj}. However, 
while dealing with general fluids this might not be the case. 
We decide to look at the imperfect dark energy fluid model developed and analysed by \cite{Hu:1998kj} 
and \cite{Koivisto:2005mm,Mota:2007sz}.
In this fluid approach, the anisotropic stress is a viscosity term damping density perturbations. 
This anisotropy is sourced by scalar velocity perturbations and is generated by the shear term in the metric fluctuation; 
moreover it has to be gauge invariant \cite{Hu:1998kj}. 

In this class of models the dark energy anisotropic stress $\sigma$ is a 
function of scale $k$ and time $a$.
The equation for $\sigma$ is given, in Newtonian gauge (see \cite{Hu:1998kj}), by 
\be
\sigma' + \frac{3}{a} \sigma = \frac{8}{3} \frac{\cv}{(1+w)^2} \frac{V}{a^2H} \label{eq:sig}
\ee
where the viscosity parameter $\cv$ quantifies the coupling of the anisotropic 
stress to the velocity perturbation. On the left hand side of Eq.~(\ref{eq:sig}) also appears 
a Hubble drag term. 
This equation, governing the evolution of the anisotropic stress, is built to recover the free 
streaming equations of motion for radiation up to the quadrupole.
We remark that this ansatz does not cover mainly models in which stress fluctuations are 
not sourced by density and velocity perturbations but act as external sources for the perturbations 
\cite{Hu:1996yt}: here also vector and tensor stresses have to be modelled in order to 
understand the evolution of perturbations. 

\section{Solutions for the dark energy perturbations} \label{sec:pert}

If the universe is matter dominated, then 
$k^2\phi \simeq -4\pi G a^2 \rho_m\Delta_m \simeq -3/2\,H_0^2 \Omega_{m0} \delta_0 \equiv -\phi_0$ 
is constant and the dark matter density contrast 
grows linearly with the scale factor. Along the lines of \cite{Sapone:2009mb},
we look for solutions for $\delta$ and $V$ in presence of an anisotropic stress evolving as in 
Eq.(\ref{eq:sig}) in two regimes: that of perturbations below and of those 
above the sound horizon.

\begin{figure}
\epsfig{figure=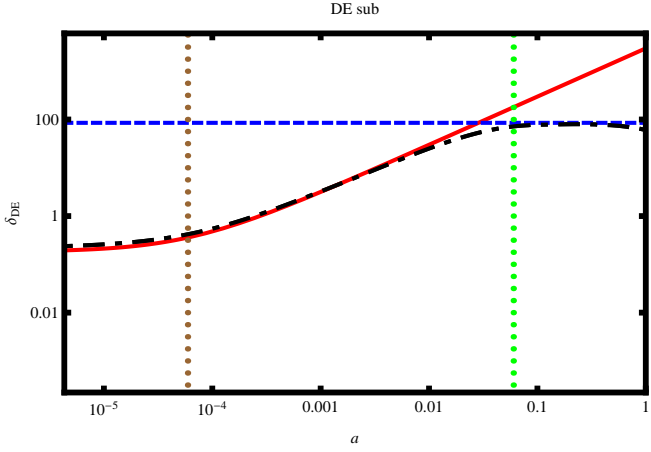,width=3.4in}
\caption{Behaviour of the dark energy density contrast $\delta$ for scales above sound 
horizon as a function of the scale factor $a$. 
The black dot-dashed line represents the numerical solution computed with CAMB for a 
model with $\cv=10^{-4}$, $\cs=0$ and $w=-0.8$ for the mode $k=200 H_0$ 
(where  radiation was omitted in order to obtain a longer interval in $a$ and show all different dynamical regimes). 
The red solid line is the approximated solution for $\cv = 0$ for scales above sound horizon, Eq.~(\ref{eq:delta0}), 
while the blue dashed line is Eq.~(\ref{eq:delta-abovesh}), valid for $\cv \neq 0$ and scales above sound horizon. 
The vertical dotted lines correspond to the value of $a$ at which the mode enters 
the causal horizon (left brown line) and the {\em anisotropic horizon} (right green line). 
We see that the numerical solution consistently follows first the red solid line and then the 
blue dashed line once the anisotropic horizon is crossed. At $a\simeq 1$ we start to see a 
deviation from our approximated solutions due to the end of matter domination, hence of their domain of validity.}
\label{fig:deltas-cs0}
\end{figure}

\subsection{Modes below sound horizon}

We start with perturbations below the sound horizon, i.e. such that $k\ll aH/c_s$. Here, 
we can neglect the terms proportional to $V$ and $V'$ appearing in Eq. (\ref{eq:vp}) as they are small compared to 
the third term on the right hand side, which is $\sim k^2 \css$. We are left with an equation for $\delta$:
\be 
\delta = \frac{1+w}{k^2 c_s^2}\left(\phi_0 + k^2 \sigma \right)\,,
\label{eq:delbelow}
\ee
where we used $k^2\,\phi=-\phi_0$. 
Substituting Eq.~(\ref{eq:delbelow}) into Eq.~(\ref{eq:deltap}) we obtain 
another differential equation for $\sigma$, coupled to $V$:
\be 
\frac{1+w}{c_s^2} \sigma' = -\frac{V}{a^2 H} - \frac{3}{a}\left(c_s^2 -w \right) \frac{1+w}{k^2 c_s^2} \left(\phi_0+k^2 \sigma\right) \,
\label{eq:sigprime}
\ee
where we have neglected the second term in square brackets of Eq. (\ref{eq:deltap}), as $k^{-2}$ is small for modes below sound horizon. Inserting the expression for $V$ of Eq. (\ref{eq:sigprime}) into Eq. (\ref{eq:sig}),  we obtain a first order differential equation in $\sigma$ only
\bea
&&\left[ \frac{1}{\css} + \frac{3}{8} \frac{1+w}{ \cv} \right] \sigma' + \frac{3}{a}\left[ \frac{3}{8} \frac{1+w}{\cv} + \frac{\css -w}{\css} \right] \sigma =\nonumber \\
&&= -\frac{3}{a} \frac{\css -w}{k^2 \css}\phi_0 \,.
\eea
This equation can be solved exactly, the solution being
\be
\sigma = - \frac{8 \cv \left( \css -w\right)}{3 \css(1+w) + 8 (\css -w) \cv} \frac{\phi_0}{k^2}\,,
\label{eq:sigma-sub}
\ee
plus an extra term rapidly decaying when $a$ grows (see Appendix \ref{sec.appx}).
Substituting  $\sigma$ into Eqs. (\ref{eq:delbelow}) and (\ref{eq:sigprime}) we finally obtain expressions for   $\delta$ and $V$:
\bea
\delta &=& \frac{3(1+w)^2}{3 \css (1+w) + 8 \left( \css -w \right) \cv} \frac{\phi_0}{k^2}  \,,  \label{eq:delta-sub-below}\\
V &=& - \frac{9(1+w)^2 \left( \css -w \right)}{3 \css (1+w) + 8 \cv (\css -w)} H_0 \sqrt{\Omega_m} \frac{\phi_0}{\sqrt{a}k^2}  \nonumber \\
&=& -3  a H \left(\cs -w \right) \delta\,. \label{eq:V-sub-below}
\eea
We notice that in our approximations $\sigma$ and $\delta$ are approximately constant while $V \sim a^{-1/2}$.

Dark energy perturbations in this scenario are suppressed by a factor $(1+w)^2$. 
The $(1+w)^2$ factor is originated both by the  $(1+w)$ factor contained in the gravitational potentials, 
which source dark energy perturbations, and by the anisotropic stress. Therefore, 
anisotropic stress slows the growth of dark energy perturbations by an extra factor of $(1+w)$ with 
respect to the standard case of clustering quintessence. 
In the limit of no viscosity,  $\cv = 0$, we consistently obtain $\sigma = 0$ and
\bea
\delta &=& (1+w)\frac{\phi_0}{\css k^2} \,,\\
V &=& - \frac{3 (1+w) \left(\css -w \right) H_0 \sqrt{\Om}}{\css k^2} a^{-1/2} \,,
\eea
which corresponds to the standard case studied in \cite{Sapone:2009mb}. Moreover 
we see that now the suppression factor of dark energy perturbations is again $(1+w)$ and not anymore $(1+w)^2$.

\subsection{Modes above sound horizon}

To find solutions for $\sigma$, $\delta$ and $V$ for modes above sound horizon,
we must again start from the velocity perturbation equation (\ref{eq:vp}) but this time we neglect the term 
proportional to $ \css$ so we have
\be
V' + \frac{1}{a} V = \frac{(1+w)}{a^2 H}\left(\phi_0 +k^2 \sigma\right)\,.
\label{eq:vbelow}
\ee
This equation, together with Eq. (\ref{eq:sig}), calculated in the approximation of matter domination i.e. for $a^2H = H_0^2 \sqrt{\Om}\sqrt{a}$, 
forms a closed system that can be solved after some mathematical manipulation, 
giving the following solution for $\sigma$
\be
\sigma = \frac{1}{k^2} \phi_0 \left[ -1 + \frac{3}{2a \alpha} +  \frac{3}{2(a \alpha)^2} +  \frac{9}{4(a \alpha)^3} \right]\,, 
\label{eq:sigma-sub-cs0}
\ee
where
\be\label{eq.alpha}
\alpha \equiv \frac{8 \cv}{3 H_0^2 \Omega_m} \frac{k^2}{1+w}\,,
\ee
plus some extra decaying terms which we can neglect.
As in the case of modes below sound horizon, the dominant term is a constant. 
Looking again at Eq. (\ref{eq:vbelow}), we now notice that, 
contrary to the standard case where $\cv = 0$, here too we could have neglected the terms 
$\propto V$, $V'$, since they decay as $a^{-1/2}$ compared to $\sigma$. 
Indeed neglecting them, Eq. (\ref{eq:vbelow}) gives us consistently $\sigma = -\phi_0/k^2$. 
This also coincides with the solution for $\sigma$  we found for modes below sound horizon 
(where we also had neglected terms  proportional to $ V$, $V'$), if we set $\css = 0$ 
into our solution, and the same happens to $V$, which is
\be \label{eq:V-sub-above}
V = -\frac{9(1+w)^2}{8 \cv}H_0 \sqrt{\Om} \frac{\phi_0}{k^2} a^{-1/2}\,.
\ee
We compute $\delta$ from Eq. (\ref{eq:deltap}), where we neglect terms $\sim \css$ and $\sim k^{-2}$:
\be
\delta' -\frac{3w}{a} \delta = -\frac{9 (1+w)^2}{8 \cv a} \frac{\phi_0}{k^2} 
\ee
and the (dominant term in the) solution is
\be \label{eq:delta-abovesh}
\delta = -\frac{3(1+w)^2}{8 \cv w} \frac{\phi_0}{k^2}\,.
\ee
It is clear again that the last equation is a subcase of solution (\ref{eq:delta-sub-below}). 
Looking at Eqs.~(\ref{eq:sigma-sub-cs0}-\ref{eq.alpha}) it would seem that the terms containing $\cv$ 
may come to dominate if we are dealing with small values of $\cv$, especially at early times. 
However, the impact of the these terms on the growth of the dark energy density contrast is negligible, 
as explained below.

Let us remind that these results 
have been obtained under the assumption of a time-independent 
$w$, $\cs$ and $\cv$, but we have not excluded a $k$-dependence of $\cs$ and $\cv$.
It is also important to notice that the solutions found above do not  
reduce to the case $\cs=\cv=0$. For this case we need to use the 
solution found in \cite{Sapone:2009mb} for the density perturbation: 
\be
\delta = \delta_0(1+w)\left(\frac{a}{1-3w}+\frac{3H_0^2\Om}{k^2}\right)\,.
\label{eq:delta0}
\ee
In Fig.~\ref{fig:deltas-cs0} we plot the numerical solution for the dark energy density 
contrast for $k=200H_0$ as well as the analytic solutions for modes above sound horizon 
($\cs = 0$), Eqs. (\ref{eq:delta-abovesh}) and (\ref{eq:delta0}) 
for $\cv=10^{-4}$ and $\cv = 0$ respectively; here and in the following figures we set 
$\Om = 0.23$ and $H_0 = 70\,{\rm km/s/Mpc}$. 
The values are chosen so to show the full complexity of the $\delta$ evolution. 
It can clearly be seen how perturbations start to grow as soon as they enter the 
causal horizon. Although in this case there is no sound horizon as the sound speed 
is zero, the presence of a non-vanishing anisotropic stress creates a new {\em anisotropic horizon} 
in the evolution of the dark energy perturbations: 
when perturbations cross this horizon then the viscosity of the fluid 
counteracts the gravitational collapse and prevents the perturbations from growing. 
The value of the scale factor at which perturbations cross the anisotropic horizon 
can be evaluated from $k c_{vis}  \sim a H $\footnote{Even if this expression is only a qualitative 
estimate of the {\em anisotropic horizon}, it indicates well the time when perturbations start to 
feel the viscosity of the fluid (see Figs. \ref{fig:deltas-cs0} and \ref{fig:deltas-cs001-cvs}). 
The reason of this is explained later in this section}.

It is the presence of this anisotropic horizon that justifies the assumptions made so far. 
To illustrate this, let us consider again the solution for $\sigma$ obtained assuming $\cs=0$, i.e. Eq~(\ref{eq:sigma-sub-cs0}).
If the sound speed is zero then the only component capable to prevent dark energy perturbations from growing 
is the anisotropic stress, which starts becoming important and counterbalancing the gravitational 
collapse only after the anisotropic horizon has been crossed by $\delta$. A way to understand this is 
by looking at Eq. (\ref{eq:sig}): when $\cv$ is very small, the evolution of $\sigma$ is essentially 
decoupled from $V$, hence the term proportional to $\sigma$ in Eq. (\ref{eq:vbelow}) is negligible 
and the anisotropic stress does not affect the growth of perturbations. Only later in time, 
approximately when crossing the anisotropic horizon, does viscosity enter the game. 

As shown, the size of this horizon depends on the value of the viscosity term $\cv$. During matter domination, 
the anisotropic horizon crossing happens at $a \sim H_0^2 \Om/(k^2 \cv)$: the smaller 
is $\cv$, the later in time perturbations enter the horizon and consequently 
they have more time to grow. The reader might think that the terms appearing in 
Eq.~(\ref{eq:sigma-sub}) come to dominate eventually for a small value of $\cv$, when $a$ is small. 
However, this is not the case thanks to the presence of the anisotropic horizon. 
This is illustrated in Fig.~\ref{fig:deltas-cs0} where $\cs=0$ and $\cv$ assumes a very small value,
$\cv=10^{-4}$ . Before entering the anisotropic horizon 
dark energy perturbations grow as if the anisotropic stress were absent, following Eq.~(\ref{eq:delta0}).
Only when the anisotropic horizon has been crossed perturbations stop growing 
because $\sigma$ becomes important. However, this only happens at late times, as $\cv$ is small. 
Hence, being $a$ large, only the first constant term in Eq.~(\ref{eq:sigma-sub-cs0}) is important. 
To prove this, let us consider the decaying mode in Eq.~(\ref{eq:sigma-sub-cs0}) for $\cv=10^{-4}$, 
\be\label{eq:decaying}
\frac{1}{\alpha a} = \frac{3}{8}\frac{H_0^2\Om}{\cv k^2}\left(1+w\right)\frac{1}{a}\sim\frac{10^{-2}}{a}
\ee
where $k=200H_0$. The above term is larger than the constant term if $a<10^{-2}$, but it can never 
dominate because the anisotropic horizon is at $a\sim10^{-1}$.

What happens at large scales when $\cv$ is larger, say $\cv \sim 10^{-1}$? From Eq. (\ref{eq:sigma-sub-cs0}) 
it would seem again that $\sigma$ cannot be approximated by a constant. Here a different mechanism 
comes into play: large scales enter the causal horizon later than small scales (e.g. during matter 
domination at $a \sim H_0^2 \Om /k^2$), hence perturbations with small $k$ cross the causal 
horizon at sufficiently large $a$ so that the time-dependent terms in Eq. (\ref{eq:sigma-sub-cs0}) 
can be neglected. If e.g. we take scales as large as $k=H_0$, the causal horizon is at $a = \Om$ 
and here we get (again for $w = -0.8$)
\be
\frac{1}{\alpha a} = \frac{3}{8} \frac{(1+w)}{\cv } \sim 0.2 < 1.
\ee
This quantity becomes even smaller for values of $w$ closer to $-1$.  The considerations made in this 
section ensure us that the time dependent terms (\ref{eq:decaying}) decay fast enough in all relevant cases.

\subsection{An effective sound horizon}

\begin{figure}
\epsfig{figure=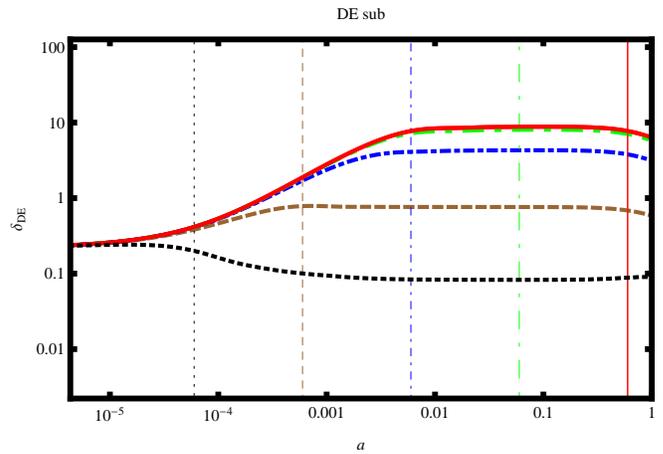,width=3.4in}
\caption{Numerical evolution of the dark energy density perturbation $\delta$ with $k=200H_0$ for 
different values of $\cv$. 
The black dotted, the brown long-dashed, the blue dot-dashed, the green long-dash-dotted and 
the red solid line (bottom to top) correspond to $\cv=10^{-1}, \,10^{-2},\,10^{-3},\,10^{-4}$ and $10^{-5}$, 
respectively. The vertical lines give the scale factor at which each mode of matching colour/line style enters the {\em anisotropic horizon} 
(left to right). For all solutions, $w=-0.8$, $\cs=10^{-2}$ was assumed.
}\label{fig:deltas-cs001-cvs}
\end{figure}

It is interesting to notice that the anisotropic stress is not always the dominant component 
affecting the growth of dark energy perturbations. As an example, we plot in Fig.~(\ref{fig:deltas-cs001-cvs}) 
the evolution of the dark energy density contrast $\delta$ for different values of the viscosity term $\cv$ 
(with the corresponding anisotropic horizon) while keeping the sound speed $\cs$ fixed. 
As the viscosity term decreases, perturbations start to feel the presence of pressure perturbations, which, 
similarly to $\cv$, damp $\delta$ once it crosses the sound horizon.
At this point, having density perturbations already been damped, the anisotropic horizon plays no role 
anymore and the viscosity term becomes unimportant. This effect can be clearly seen in 
Fig.~(\ref{fig:deltas-cs001-cvs}), where the red solid line and the green long-dash-dotted line, 
corresponding to $\cv = 10^{-5}$ and $\cv=10^{-4}$, respectively, are basically unaltered when 
crossing their anisotropic horizon (that for the value used it happens at $a\sim 10^{-3}$), 
having already experienced the damping due to pressure perturbations. 
A similar behaviour can be seen when the anisotropic horizon is crossed before the sound horizon: 
here $\delta$ is damped by the viscosity of the fluid and the crossing of the sound horizon 
later in time produces no effect. 
Furthermore, looking at Eqs. (\ref{eq:delta-sub-below}-\ref{eq:V-sub-below}) we notice that it is 
possible to rewrite them in terms of an {\em effective sound speed}:
\be
\ceff = \cs +\frac{8}{3} \frac{(\cs -w )}{(1+w)} \cv\,.
\ee
We list the equations for $\delta$ and $V$ in terms of $\ceff$ and $\cs$ in Appendix \ref{sec.appz-eff-cs}. 
In view of these considerations, the degeneracy between $\cs$ and $\cv$ found by us, 
\cite{Koivisto:2005mm} and  \cite{Calabrese:2010uf} is confirmed and has a clear explanation.
From $\ceff$ it is hence possible to define an {\em effective sound horizon}, which takes into 
account both $\cv$ and $\cs$ and determines the time at which perturbations are damped by 
the anisotropic stress or by pressure perturbations: $a^2 H^2 \sim k^2[\cs+8/3 \cv(\cs-w)/(1+w)]$.
We can ask when the anisotropic stress comes to dominate over pressure perturbations. 
The sound speed starts becoming more important than anisotropic stress if $\cv<\cs/10$ 
(for our value of $w$), which is also confirmed by numerical results.

\section{Clustering parameters} \label{sec.QS}

When dark energy starts dominating, $\phi$ starts decaying; moreover,
dark energy is also clustering. 
This variation can be conveniently represented using the parameterisation 
defined in \cite{Amendola:2007rr} where two new 
quantities were introduced: $Q$, which is directly related to the Poisson equation and describes the amount of 
clustering dark energy, and $\eta$, which is related 
to the amount of anisotropic stress. 
Using this new parameterisation, the Poisson equation reads
\be
k^2 \phi = -4 \pi G a^2 Q \rho_m \Delta_m
\label{eq:phi-q}
\ee
where 
\be \label{eq.Q}
 Q -1= \frac{\rho \Delta}{\rho_m\Delta_m}\,.
\ee
The reason why we mostly express quantities in terms of $Q-1$ rather than $Q$ itself is because the former 
directly measures the amount of clustering dark energy whereas $Q$ refers to the sum of dark 
matter and dark energy clustering.
The parameter $\eta$ is instead defined from
\be
\psi = \left(1+\eta\left(a,k\right)\right)\phi\, .
\label{eq:psi-eta}
\ee
We can now compute the gauge-invariant density perturbation $\Delta$ of matter and 
dark energy. During matter domination we have at first approximation $\Delta_m = \delta_0\,a$, 
where $\delta_0$ is a constant of integration.
For dark energy we obtain
\be
\Delta = \frac{(1+w)}{\ceff} \frac{\phi_0}{k^2}
\label{eq:delta-sub}
\ee
where we have neglected the term proportional to $V$ since it is decaying. 
Also recall that this solution is valid for modes both above and below 
sound horizon. 

In our case, we will find an expression for $Q-1$ valid for all $\cv \geq 0$ (and $\cs\geq 0$):
\be
Q-1=\frac{3H_0^2}{2k^2}\frac{\left(1-\Om\right)(1+w) a^{-3w-1}}{\ceff}\,.
\label{eq:Q-above}
\ee
We notice that the above equation is not valid for the special case $\css = \cv = 0$ (as it 
was already the case for $\delta$). Being the latter the standard case with no anisotropic stress, 
it had already been calculated in \cite{Sapone:2009mb} to be
\be
Q({\css=\cv=0})-1 = \frac{1+w}{1-3w}\frac{1-\Om}{\Om} a^{-3w}\,.
\ee
To obtain a single expression for $Q$, including anisotropic stress as well as the case $\cs=\cv=0$ 
and valid for modes above and below sound horizon, we define
\bea \nonumber
&Q_{tot}&-1\equiv \frac{\left[Q -1\right] \left[Q({\css=\cv=0})-1\right] }{ \left[Q -1\right]+ \left[Q({\css=\cv=0})-1\right] }\\
&=& \frac{1-\Om}{\Om}(1+w)\frac{a^{-3w}}{1-3w + \frac{2 k^2 a}{3H_0^2\Om}\ceff}\,.
\label{eq:qtot}
\eea
The above expression is extremely simple and it looks like the solution found in 
\cite{Sapone:2009mb}, except for the term including the effective sound speed $\ceff$ 
which depends both on the sound speed and on the viscosity term.
It is easy to see that if the anisotropic stress is absent then Eq.~(\ref{eq:qtot}) reduces to 
the expression found in \cite{Sapone:2009mb}. From Eq. (\ref{eq:qtot}) we have a further 
confirmation of the degeneracy of $\cs$ and $\cv$, when looking at quantities which depend 
only on the clustering of dark energy. To break this degeneracy, we will have to combine 
measurements of $Q$ with measurements of $\eta$, defined in Eq.~(\ref{eq:psi-eta}), as will be clear from the following.

\begin{figure}
\epsfig{figure=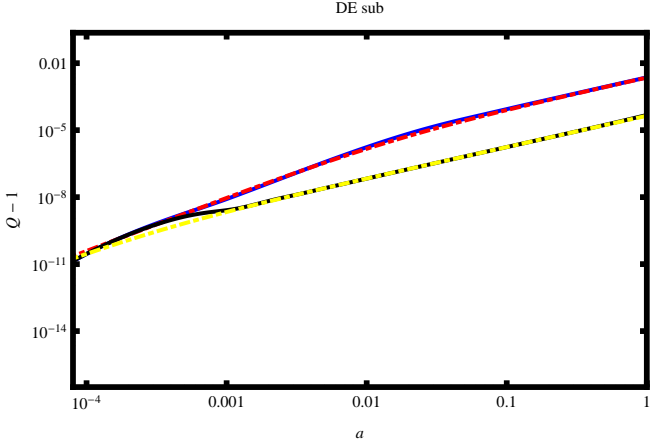,width=3.4in}
\caption{Behaviour of $Q_{tot}-1$ for 
two different values of the viscosity term $\cv$. 
The solid lines are numerical solutions computed with CAMB (blue/upper line: $\cv = 10^{-4}$, black/lower line: $\cv=10^{-2}$) 
while the red dashed line and the yellow dot-dashed line represent Eq.~(\ref{eq:qtot}) with  $\cv = 10^{-4}$ 
and  $\cv = 10^{-2}$, respectively. All lines correspond to $k=200H_0$, $w=-0.8$ and  $\cs=10^{-3}$.
}\label{fig:qtot}
\end{figure}

Let us indeed evaluate the quantity $\eta$:
\bea
&\eta& = -\frac{k^2\left(\phi-\psi\right)}{k^2\phi}=\frac{B\left(a\right)}{\phi_0\,Q_{tot}}\sigma \nonumber \\
&=& -\frac{9}{2}H_0^2(1-\Om)(1+w)\frac{a^{-1-3w}}{k^2 Q_{tot}}\left(1- \frac{\cs}{\ceff} \right)\,
\label{eq:eta-our}
\eea
where we used the relation $k^2\phi= -\phi_0Q_{tot}$. This parameter depends not only on 
$\ceff$ but on the ratio of the sound speed and $\ceff$. For this reason, the only way of 
breaking the degeneracy between $\cv$ and $\cs$ (at least at the linear perturbation level) 
is through the combination of observables constraining $Q$ and $\eta$. The detection of a 
non-zero $\eta$ alone would prove the existence of an anisotropy but it would not allow to 
quantify it as only the ratio $\cs/\ceff$ would  be measurable.

If instead of  $\eta$ we prefer to use the variable $\Sigma$ introduced in \cite{Amendola:2007rr} 
and more strictly related to the weak lensing potential $\Psi = \phi+\psi$, which is defined as
\be
\Sigma = Q\left(1+\frac{1}{2}\eta\right)\,,
\ee
we can evaluate it analytically using Eqs.~(\ref{eq:qtot}) 
and (\ref{eq:eta-our}): 
\bea
&&\Sigma_{tot}-1=Q_{tot}\left(1+\frac{\eta}{2} \right)-1 =\nonumber \\
&&=\left(Q_{tot}-1\right)\left[1-\frac{3}{2}\left(1-\frac{\cs}{\ceff}\right)\frac{1+\beta a}{\beta a}\right]
\label{eq:Stot}
\eea
where $\beta = 2k^2\ceff/[3H_0^2\Om(1-3w)]$. 
The above equation shows that the weak lensing parameter $\Sigma$ differs from 
$Q$ by about $10-15\%$ (depending on $a$) in case of a non-zero anisotropic stress. 

\begin{figure}
\epsfig{figure=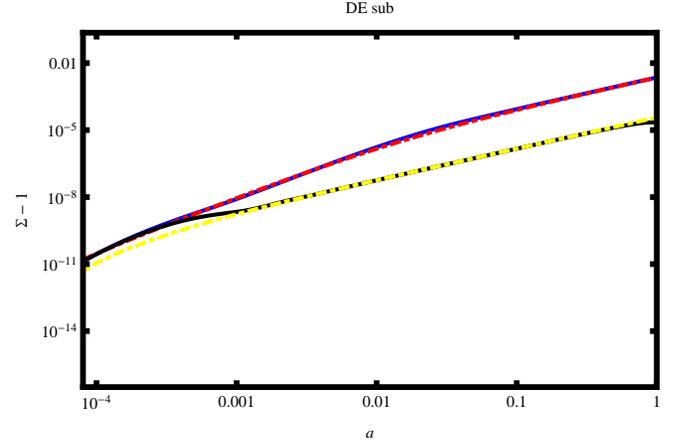,width=3.4in}
\caption{Behaviour of $\Sigma_{tot}-1$ for 
two different values of the viscosity term $\cv$. 
The solid lines are CAMB outputs (blue/upper line: $\cv = 10^{-4}$, black/lower line: $\cv=10^{-2}$) 
while the red dashed line and the yellow dot-dashed line represent Eq.~(\ref{eq:qtot}) with  
$\cv = 10^{-4}$ and  $\cv = 10^{-2}$, respectively. All lines correspond to $k=200H_0$, $w=-0.8$ and 
$\cs=10^{-3}$.}\label{fig:Stot}
\end{figure}

In Figs.~\ref{fig:qtot} and \ref{fig:Stot} we compare $Q_{tot}$ and $\Sigma_{tot}$, respectively, 
with their numerical solution, computed using CAMB, for different values of the dark energy viscosity term 
keeping the sound speed fixed. We use two different values of $\cv$ that are respectively 
ten times smaller and ten times larger than $\cs$;  these values were chosen
such that in the first case the sound speed is the dominant component while in the second the 
viscosity term dominates. We find very good agreement between the numerical solution and 
our analytical estimate, as it was the case in~\cite{Sapone:2009mb}.

\section{Effect on some observables} \label{sec:obs}

\subsection{Matter power spectrum} \label{sec:matterPk}

As in \cite{Sapone:2009mb}, after evaluating the impact of anisotropic stress on dark energy 
perturbations, we move to look at the (much smaller) effect it should have on matter perturbations. 
These are affected by the change in gravitational potential from $\phi_m =  -3/2 \delta_0 H_0 \Om/k^2$ to 
$\phi = \phi_m +\phi_{\rm DE}$, where $\phi_{\rm DE} = \phi_m (Q-1)$. 
For modes larger than the dark energy sound horizon, $Q$ is given by 
Eq. (\ref{eq:Q-above}) where $\cs=0$ and we can solve for the 
dark matter velocity perturbation, Eq. (\ref{v}) with $\delta p_m = 0$,  to find
\be \label{eq:Vmatter}
V_m = -H_0 \sqrt{\Om} \delta_0 \sqrt{a} \left[1- \frac{Q_0 a^{-3w -1}}{1/3-2w} \right],
\ee
where 
\be
Q_0 = \frac{9}{16} \frac{H_0^2 }{k^2} (1-\Om) \frac{(1+w)^2}{w}\frac{1}{\cv}
\ee
and where the decaying term has been omitted.
This equation is of course strictly valid only during matter domination, but it gives again the 
right order of magnitude of the effect even after this epoch, as we will see in the following. 
The correction factor with respect to the case of no dark energy perturbations is the term in square brackets. 
We notice that its evolution in time, contained in the second term in square brackets 
and proportional to $a^{-3w -1}$, is different from that in absence of anisotropic stress, 
which was  $\sim a^{-3w}$. 
Substituting Eq. (\ref{eq:Vmatter}) into Eq. (\ref{delta}) (clearly with $w_m= \delta p_m = 0$) 
we obtain a differential equation for $\delta_m$, whose solution is
\bea \nonumber
\delta_m &=&  \delta_0 \left\{ a\left[1+ \frac{Q_0 a^{-3w -1}}{w(6 w -1)} \right] + \right. \\
&& + \left. \frac{3 H_0^2 \Om}{k^2} \left[ 1 + \frac{3}{2} Q_0 a^{-3 w -1} \right] \right\}. \label{eq:deltam-cv}
\eea
As in the case analysed by \cite{Sapone:2009mb}, the enhancement factor due to the viscosity 
term varies depending whether perturbations are inside or outside the causal horizon. 
In the first case and for perturbations outside the sound horizon, the first term in curly 
brackets is more important so that the correction term is $Q_0 a^{-3w}/[w(6 w -1)]$, 
while for perturbations outside the causal horizon, the term is $ 9 H_0^2 \Om  Q_0 a^{-3 w -1}/(2k^2)$. 

\begin{figure}
\epsfig{figure=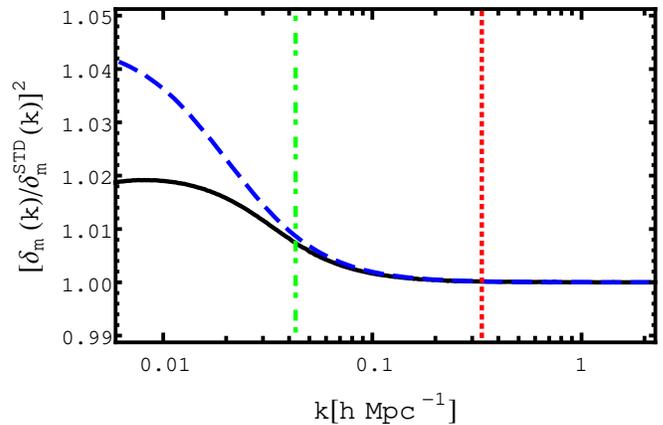,width=3.4in}
\caption{Behaviour with $k$ of the ratio of $\delta_m^2$ to the square of the standard matter perturbation 
in absence of dark energy perturbations $[\delta_m^{\rm STD}]^2$. The black solid curve 
represents the numerical solution found using CAMB with $\cv = 5 \times 10^{-5}$ and 
$\cs = 10^{-6}$. The blue dashed curve is our analytical solution, computed from Eq. (\ref{eq:deltam-cv}). 
The green dot-dashed vertical line represents the anisotropic horizon while the red dotted line is the sound horizon.
Here we fix $w= -0.8$ and $a = 1$.}
\label{fig:deltam}
\end{figure}

Let us now compare the matter power spectrum $P(k, a) \sim \delta_m(k, a)$ obtained in the 
case of $\cv \neq 0$  to the standard one, obtained in absence of dark energy perturbations 
$P^{\rm STD}(k) \sim \delta_m^{\rm STD}(k, a)$. In Fig. \ref{fig:deltam} we plot the ratio of 
these two quantities. As can be seen, our approximation of $\delta_m$ is of the same order 
of magnitude of the numerical solution, which is a good result for a second order quantity 
such as $\delta_m$. We also see that once the matter perturbations enter the anisotropic horizon, 
the analytical and numerical solution coincide and both tend to the solution of unclustered dark energy. 
This is consistent with what we have learned from the previous section, i.e. that dark energy 
perturbations inside the anisotropic (or effective sound) horizon are damped and can be neglected. 
An implication of this is that anisotropic stress will be difficult to detect being present 
only at large scales. Also, for $w=-0.8$ we expect a maximum $4\%$ enhancement of 
$P(k)$ on scales larger than both the anisotropic and the sound horizon. Our numerical calculation shows that the enhancement
is closer to $2\%$, hence the effect is very small. 

Having evaluated the impact of the anisotropy on the matter power spectrum amplitude, we move to compute explicitly a more easily observable parameter.

\subsection{The growth factor}

In the $\Lambda$CDM scenario, dark matter perturbations 
grow logarithmically with the scale factor $a$ during 
radiation domination and when matter domination starts they grow linearly; 
at late times, when dark energy starts to dominate, the growth of dark matter perturbations 
is suppressed. It is known that in the $\Lambda$CDM model the growth
factor can be defined as: 
\be
G\left(a\right)\equiv\frac{\delta_{m}(a)}{\delta_{m}(a_{0})}=\exp\Big\{\int_{0}^{a}\frac{\Omega_{m}\left(a'\right)^{\gamma}}{a'}{\rm d}a'\Big\}
\ee
where $\gamma\sim0.545$ is called the growth index. There are two
ways to modify the growth factor. First, one can use a different background cosmology, 
associated to a different Hubble expansion. Second, perturbations can differ:
if dark energy starts to cluster, then the gravitational potentials change, being sourced 
by the total density perturbations and by the anisotropic stress
(see the Poisson equation (\ref{eq.Poisson_general}) and the fourth Einstein equation (\ref{eq.4thEinsteinB})) 
and this will affect the growth rate of dark matter. 
It is possible to include all these effects in the growth index
$\gamma$ and we therefore expect $\gamma$ to be a function of
$w$, $\cs$ and $\cv$(or equivalently of $w$, $Q$ and $\eta$).

In \cite{lica}, the dependence of the growth index from the dark energy clustering parameters $Q$ and $\eta$ was described as
\be
\gamma=\frac{3\left(1-w-A\left(Q,\eta\right)\right)}{5-6w}
\label{eq:gamma-Qeta}
\ee
where 
\be
A\left(Q,\eta\right)=\frac{\left(1+\eta\right)Q-1}{1-\Omega_{m}\left(a\right)}\,.
\label{eq:A-Qeta}
\ee
While dark energy perturbations themselves are difficult to
measure unless  dark energy has very low sound speed and viscosity, the
growth index seems to be a more easily detectable parameter and several ongoing and future
experiments are built to measure its value. 
It was shown in \cite{Sapone:2010uy} that the presence 
of a sound speed (when $\cv=\eta=0$) in the dark energy perturbations 
always decreases the value of the growth index $\gamma$ 
because $Q-1$ is always positive due to the relative increase of 
dark energy perturbations. 
Here instead the addition of a non-zero 
anisotropic stress may lead to an {\em enhancement} of the 
growth index. To prove this let us consider a particular case. 
If we take modes above the 
sound horizon, using Eqs.~(\ref{eq:Q-above}) and Eq.~(\ref{eq:eta-our}) with 
the sound speed set to zero, Eq.~(\ref{eq:A-Qeta}) reads 
\be
A\left(Q,\eta\right) = B(a)\frac{1}{k^2}\left(\frac{1+w}{-8w\cv}-1\right)\,.
\ee
The function $B(a)$ is always positive as long as $w$ is larger than $-1$, hence the 
term in brackets can be negative only if $\cv>(1+w)/(-8w)$. 
If we assume $w=-0.8$ then $A(Q,\eta)$ is negative as long as $\cv>1/32$ and $\gamma$ is enhanced. 
However, the effect of the anisotropic stress on scales of interest is very small 
due the smallness of the factor $B(a)/k^2$ in Eq.~(\ref{eq:A-Qeta}) that contains the term $(1+w)/k^2$. 

Again, if the anisotropic stress is zero, then
dark energy perturbations always decrease the value of the growth
index; we find that the strongest deviation from $\gamma_{\Lambda CDM} = 0.545$ happens 
when $\cs\simeq 0$ and $\cv \simeq 0$ and it is of the order of $3$\%, see \cite{Sapone:2009mb}. 
If we want to increase the growth index and at the same time we also want 
perturbations in the dark energy sector then we need a
non zero anisotropic contribution; this is the case for the DGP model
(treated as an effective anisotropic dark energy, see \cite{Kunz:2006ca})
where the growth index is $\gamma\sim0.68$.
However, the type of anisotropic stress used in this paper is unable to increase 
the growth rate $\gamma$ from $\gamma_{\Lambda CDM}$ to such a large value. For instance, 
if we assume the viscosity term to be $\cv=1$ then $A\left(Q,\eta\right) \simeq -1.5\times10^{-5}$ for 
scales $k\simeq 200H_0$.

\subsection{The Integrated Sachs-Wolfe effect}

Another potentially observable effect of anisotropies is the integrated Sachs-Wolfe (ISW) effect. 
Indeed, as explained e.g. in \cite{wellew,beandore,bms,ddw}, being this a late-time effect, 
it is the part of the CMB spectrum where dark energy has the largest impact.
The total temperature shift of the CMB photons due to the ISW is \cite{Sachs-Wolfe}:
\bea
\zeta &=& \frac{\Delta T\left(\hat{n}\right)}{T_{0}}=\int{\left(\frac{\partial\phi}{\partial\tau}+\frac{\partial\psi}{\partial\tau}\right)}d\tau=\nonumber\\
&-&\int_{0}^{\chi_{_H}}{a^{2}H\frac{\partial\Phi}{\partial a}}d\chi, \label{eq:ISW1}
\eea
where $\chi$ is the comoving distance ($d\chi =-cd\tau=-cdt/a$). 
In this work we decide to consider the effect of the anisotropic 
stress only for dark energy, in order to isolate its contribution and to compute its size analytically. 
See \cite{Calabrese:2010uf} for a numerical analysis which includes massive neutrinos and their anisotropic stress.
The term inside the integral is the derivative of the weak lensing potential $\Phi=\psi+\phi$
with respect to the scale factor, which in Fourier space is
\bea
&&\Phi'=-\frac{3}{2} \frac{H_{0}^{2}\Om}{ak^2} \left\{\Sigma\left(a,k\right)\Delta'_{m}\left(a,k\right) + \right. \nonumber \\
&&+\left. \Sigma'\left(a,k\right)\Delta_{m}\left(a,k\right)-\frac{1}{a}\Sigma\left(a,k\right)\Delta_{m}\left(a,k\right)\right\}. 
\label{psidot}
\eea
From this expression we can see that anisotropic dark energy perturbations modify the ISW effect 
through the changes induced in $\Delta_m$ and through the additional presence of $\Sigma$ and $\Sigma'$.

\begin{figure}
\epsfig{figure=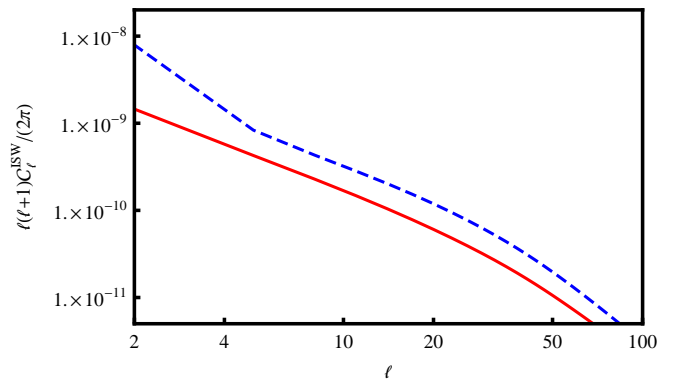,  width=3.4in}
\caption{ISW power spectrum for $\cv= 0$ (red solid line)
and $\cv =10^{-2}$(blue dashed line) fixing $w=-0.8$ and $\cs=10^{-4}$.}
\label{fig:ISW-Cls}
\end{figure}

At linear order, it is possible to isolate today's $\Delta_m$ from its time evolution: 
\be
\Delta_{m}\lakr=aG\lakr\Delta_{m,0}\lkr ,
\ee
where $\Delta_{m,0}\lkr\equiv \Delta_m(a=1,k)$.
Already in \cite{Sapone:2009mb} and  \cite{ddp} a deviation from the usual scale-independent growth factor $G(k)$ 
due to dark energy perturbations was evidenced. There, the $k$-dependence was determined by the 
presence of the sound horizon. In our case, the mechanism is the same, but depends on the {\em effective} 
sound horizon, which is determined by both the sound speed and the viscosity of dark energy.
Here we express in first approximation $G\lakr = \Omega_m(a)^{\gamma\lakr}$, with $\gamma(a,k)$ from 
Eqs.~(\ref{eq:gamma-Qeta}) and (\ref{eq:A-Qeta}). We write Eq.~(\ref{psidot}) as:
\be
\Phi' = -\frac{3}{2}\frac{H_{0}^{2}\Om}{k^2}\frac{\partial}{\partial a}\Big \{G\lakr \Sigma\lakr \Big\}\Delta_{m,0}\lkr\,. 
\label{eq:psiprimeG}
\ee
Eq. (\ref{eq:ISW1}) reads now
\bea
\zeta &=& \int_{0}^{\chi_{_H}}{d\chi W_{\zeta}\left(\chi\right)\Delta_{m,0}\lkr} \\
W_{\zeta}\left(\chi\right) &=& \frac{3}{c^3}\frac{H_{0}^{2}\Om}{k^{2}}a^{2}H\frac{\partial}{\partial a}\Big \{G\lakr \Sigma\lakr \Big\}\,.
\label{windowISW}
\eea
Hence the ISW-auto correlation spectrum $C_{\zeta\zeta}\left(\ell\right)$ is (in the Limber-projection 
\cite{Limber} and in the flat-sky approximation):
\be
C_{\zeta\zeta}\left(\ell\right)=\int_{0}^{\chi_{_H}}{d\chi\frac{W_{\zeta}^{2}\left(\chi\right)}{\chi^{2}}\bar{P}_{\Delta\Delta}\left(k =\ell/\chi\right)}
\label{clstoplot}
\ee
where $\bar{P}_{\Delta\Delta}\left(k\right)$ is the linear matter power spectrum today:
\be
\frac{k^3\bar{P}_{\Delta\Delta}\left(k\right)}{2\pi^2}=\delta^{2}_{H}\left(\frac{k}{H_0}\right)^{n+3}T^{2}\left(k\right) ,
\ee
$\delta_{H}$ is the amplitude of the present-day density fluctuations at the Hubble scale 
and $T\left(k\right)$ is the dark matter transfer function. 

In Fig.~\ref{fig:ISW-Cls} we plot $\ell\left(\ell+1\right)C_{\ell}/(2\pi)$ using 
Eq.~(\ref{clstoplot}) for two different values of the dark energy viscosity keeping the sound 
speed fixed to $10^{-4}$. 
For $T(k)$ we use the fitting formula of  \cite{ehu}, which agrees with
the numerical results from CAMB with a precision sufficient for our purposes.
In computing the matter power spectrum we neglect the effect of dark energy
perturbations evaluated in Sec. \ref{sec:matterPk}, having ascertained that it is very small. 
This means that the two curves corresponding to different $\cv$ differ because of the term $(\partial (\Sigma G)/\partial a)^2$. 

Let us then evaluate
\be
(G\Sigma)' = G'\Sigma + G\Sigma' \label{eq:sigma}.
\ee
While both $\Sigma$ and $\Sigma'$ differ from the case with no perturbations, corresponding to 
$\Sigma = 1$, $\Sigma' =0$, the first deviates too little from $1$ to explain the differences between the red and the blue
curves of Fig.~\ref{fig:ISW-Cls}. We have checked this by fixing $\Sigma=1$ and letting 
$\Sigma' \neq 0$ and noticing that the results are unchanged. This means that the effect of $\cv$ 
on the ISW comes mainly from the term $\Sigma' G$. 
Using Eq.~(\ref{eq:Stot}) and (\ref{eq:qtot}) we find
\be
\Sigma' = Q'+\frac{\sigma}{2\phi_0}B'(a)\propto \frac{1}{a}\left(\Sigma-1\right)
\ee
where the last proportionality holds because  $(Q-1)'\propto \left(Q-1\right)/a$ and $B'(a)\propto B(a)/a$ 
hence $a\Sigma'\propto (\Sigma-1)$.
Let us now go back to Eq. (\ref{eq:sigma}) and look at the final total effect.
We know that as dark energy slows down the growth of dark matter perturbations, at late times $G'$ is always negative. 
$\Sigma'$ instead can be positive or negative depending on the value of $\cv$, although in both 
cases it is a small number. Therefore we have two possibilities: the first is when
the two contributions partially cancel and so make the effect smaller (this is in general the case 
for small values of the viscosity term); the second is when the two contributions sum up enhancing the total 
effect. 
\begin{figure}
\epsfig{figure=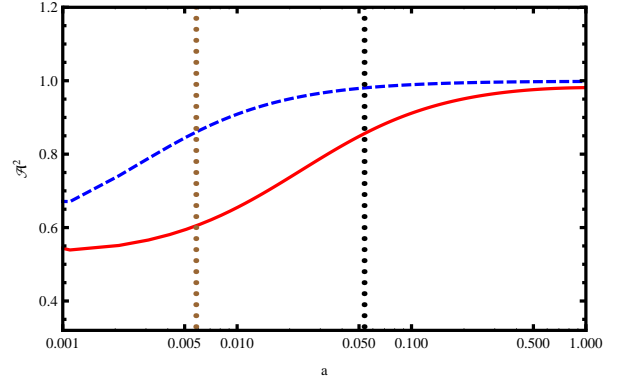, width=3.1in}
\caption{Magnification factor $\AA^2$ of Eq. (\ref{eq:magnification}) for two different values of 
the viscosity term: $\cv= 10^{-4}$ (red solid line) and $\cv= 10^{-3}$ (blue dashed line). 
Here we fix the scale of perturbations to $k=200 H_0$ and our dark energy model has $w=-0.8$ and $\cs = 10^{-4}$.
The vertical lines at $a=0.081$  ($a=0.24$) show the time at which the perturbation with 
$\cv= 10^{-3}$  ($\cv=10^{-4}$) enters the anisotropic horizon.}
\label{fig:magnification}
\end{figure}

To visualise this we proceed as in \cite{Sapone:2009mb} 
by defining a magnification parameter for the ISW power spectrum
\bea \label{eq:magnification}
\AA^2 &=& \left\{\frac{d\left(G\lakr \Sigma\lakr\right)/da}{dG\lar/da}\right\}^{2} \\
&=& (\Sigma+\Sigma' G/G')^2,
\eea
which we plot in Fig.~\ref{fig:magnification} for two values of $\cv$ for $k=200 H_0$.
Here we keep the sound speed and the viscosity term small enough so that the dark energy perturbations 
have time to grow before entering the effective sound horizons. For small $\cv$ 
the dark energy perturbations are able to grow for sufficiently long time and they partially cancel the contribution 
from $G'$ and decrease $\AA$ by about 40\%. As we increase the viscosity term $\cv$ the effects 
become smaller because the dark energy perturbations cross the effective sound horizon early, 
increasing $\AA$.  
Based on these considerations, we expect again the viscosity term to act similarly as the sound speed: 
a low value of $\cv$ decreases the power of the ISW effect, differentiating it from the standard case.

\begin{figure}
\epsfig{figure=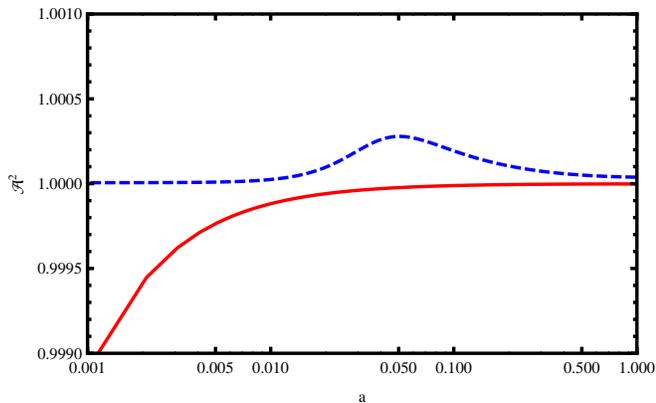, width=3.4in}
\caption{Magnification factor $\AA^2$ of Eq. (\ref{eq:magnification}) for two different values of the viscosity term: $\cv= 0$ (red solid line) and $\cv= 1$ (blue dashed line). Here we fix the scale of perturbations to $k=200 H_0$ and our dark energy model has $w=-0.8$ and $\cs = 1$ 
}
\label{fig:magnification-cv1}
\end{figure}

As mentioned before, there might be cases where $\Sigma-1$ is negative, enhancing the total ISW effect. 
In Fig.~(\ref{fig:magnification-cv1}) we show the magnification $\AA$ for $\cv=1$ and 
$\cv=0$, setting $\cs=1$ in both cases. If the sound speed is large enough then $\Sigma'$ is negative 
and it sums up to $G'$ in Eq.~(\ref{eq:sigma}). Nevertheless the effect is too small (of the order of $10^{-5}$)
and it is practically unobservable in the~$C_{\ell}$s.

\section{Conclusions}

In this paper we have studied an imperfect fluid dark energy with non-vanishing viscous anisotropic stress. 
The model we have considered is described by its background constant equation of state $w\leq0$ and at 
linear order in perturbations by its sound speed
$\cs\ge0$ and its viscosity parameter $\cv\ge 0$. For the dark energy density and velocity perturbations 
we have found simple analytical expressions represented in Eqs. (\ref{eq:delta-sub-below}) to (\ref{eq:delta-abovesh}). 
Although we have assumed matter domination throughout this work,  the functions $Q(k,a;w,\cs, \cv)$, 
$\eta(k,a;w,\cs, \cv)$ and $\Sigma(k,a;w,\cs, \cv)$ which can be computed with our simplified analytical 
expressions are approximating well numerical results even at later times, when dark energy starts dominating.
Our analytical expressions can be seen as the ``fingerprints'' of a viscous dark energy model (or of 
a modified gravity model that can be effectively described by such an anisotropic stress).

In particular we find that the dark energy perturbations are always smaller than the
perturbations in dark matter, by a factor $(1+w)$ if the viscosity term is small, $\cv\sim 0$ and by an 
even larger factor of $(1+w)^2$ when the anisotropic stress is  not negligible.

If dark energy has zero anisotropic stress then it is possible to divide the evolution of its perturbations 
in three phases: outside the causal horizon, inside the causal horizon but outside the sound horizon 
and inside the sound horizon. In the latter regime their growth is suppressed.
In the presence of a non zero viscosity term, a further horizon comes into the play, which we dubbed 
{\em anisotropic horizon}. Once perturbations enter this horizon, they are damped as in the case of the sound horizon. 
This adds in principle two extra regimes:
perturbations larger than the anisotropic horizon and perturbations smaller than the anisotropic horizon. 
The last two regimes may happen before or after the sound horizon depending if $\cv$ is larger 
or smaller than the sound speed.
In practice though, since both the sound speed and the viscosity damp the dark energy perturbations 
in a similar way, we can speak of only one {\em effective sound horizon}, which, once crossed, 
suppresses the growth of dark energy perturbations; this makes $\cs$ and $\cv$ degenerate when 
considering dark energy density perturbations.

We then have used the equations for $Q$ and $\Sigma$ to study analytically the effect of dark energy perturbations 
on the dark matter power spectrum, on the growth factor and on the ISW effect. The changes in the matter
power spectrum and in the growth index at late times are of the order of a few percent on scales larger 
than the effective sound horizon of the dark energy. 
We also have found that the  ISW effect is modified mainly by 
the time variation of the parameter $\Sigma'$ while modifications in $\Sigma$ itself do not have 
appreciable consequences. 
The parameter $\Sigma$ is a combination of the dark energy density contrast (through $Q$) and 
the anisotropic stress (through $\eta$) which both prevent perturbations from growing. Because of this, 
the two effects are difficult to capture separately as they both act similarly. 
To disentangle the two effects is possible in principle by using joint 
observations which measure $Q$ and $\Sigma$, such as e.g. the galaxy power spectrum together with the 
weak lensing maps. In a follow-up paper \cite{saponeetal}, we evaluate the 
impact of combining different experiments in order to measure at the same time the 
sound speed and the anisotropic stress of dark energy.

\section{Acknowledgments}
DS and EM thank Martin Kunz and Jon Urrestilla for useful discussions. 

DS acknowledges support from the JAEDoc program with ref. JAEDoc074 
and the Spanish MICINN under the project AYA2009-13936-C06-06.
DS also acknowledges financial support from the Madrid Regional Government 
(CAM) under the program HEPHACOS P-ESP-00346, Consolider-Ingenio 2010 PAU (CSD2007-00060), 
as well as in the European Union Marie Curie Network ``UniverseNet" under contract MRTN-CT-2006-035863.

EM was supported by the Spanish MICINNs Juan de la Cierva programme (JCI-2010-08112), 
by CICYT through the project FPA-2009 09017, by the Community of Madrid  through the project 
HEPHACOS (S2009/ESP-1473) under grant P-ESP-00346 and by the European Union ITN project (FP7-PEOPLE-2011-ITN, 
PITN-GA-2011-289442-INVISIBLES).

\appendix
\section{Decaying modes for the anisotropic stress}\label{sec.appx}

In this appendix we look carefully at the full analytic solution of $sigma$ and in particular at the 
decaying modes which we have previously neglected. 
We start by looking at modes above the sound horizon; combining Eqs.~(\ref{eq:sig}) and (\ref{eq:vbelow}) 
we obtain a second order differential equation for the dark energy anisotropic stress 
\bea
\sigma''&+&\frac{9}{2a}\sigma'+\left[\frac{3}{2a^2}+\frac{1+w}{a}\frac{k^2}{H_0^2\Om}\right]\sigma =\nonumber \\
&=&-\frac{4\cv}{1+w}\frac{\delta_0}{a^5}
\eea
being $\delta_0$ the initial dark matter density contrast that acts as a source term for dark 
energy perturbations. 
The full solution, after some mathematical manipulation, is: 
\bea
\sigma(a)&=&\frac{c_1 }{a^3\alpha^3}
\left\{\left(1-\frac{4 a \alpha}{3}\right) 
\text{Cos}\left[\sqrt{4 a \alpha}\right]+\right. \nonumber \\
&+&\left.2 \sqrt{a\alpha}\text{Sin}\left[ \sqrt{4 a \alpha}\right]\right\} \nonumber \\
&-&\frac{15 c_2}{32\,a^3 \alpha^3}
\left\{2\sqrt{a\,\alpha}  \text{Cos}\left[\sqrt{4a\alpha}\right] \right.\nonumber \\
&-&\left.\left(1-\frac{4 a \alpha}{3}\right) 
\text{Sin}\left[\sqrt{4a \alpha}\right]\right\}+\nonumber \\
&+&\frac{1}{k^2} \phi_0 \left[ -1 + \frac{3}{2a \alpha} +  \frac{3}{2(a \alpha)^2} +  \frac{9}{4(a \alpha)^3} \right]
\label{eq:full-sigma}
\eea
where $c_1$ and $c_2$ are constants of integrations depending on the initial conditions for 
$\sigma$ and 
\be
\alpha \equiv \frac{8 \cv}{3 H_0^2 \Omega_m} \frac{k^2}{1+w}\,.
\ee
Let us look more carefully: the two terms neglected in the solution Eq.~(\ref{eq:full-sigma}) are a combination 
of $\text{Sin}$ and $\text{Cos}$ which are oscillating functions if $\alpha>0$. 
Given that $\cv$ is assumed to be non-negative in this work, $\alpha$ is always positive 
as long as $-1\leq w \leq 0$. 
In case of ``phantom" dark energy instead there may be a growing solution as 
$\alpha$ may become negative. One way out of this problem is to assume 
a negative viscosity term only for phantom model, as pointed out in \cite{Mota:2007sz}.
Moreover, the oscillating terms in Eq.~(\ref{eq:full-sigma}) are also decaying modes because 
they are multiplied by the term $1/(a\alpha)^3$.

Let us now consider modes below the sound horizon, {\em i.e.} $\cs\neq 0$; combining 
Eqs.~(\ref{eq:sig}), (\ref{eq:delbelow}) and (\ref{eq:sigprime}), the 
full first order differential equation for the anisotropic stress is: 
\bea
&&\left\{ \frac{1}{\css} + \frac{3}{8} \frac{1+w}{ \cv}\left[1+\frac{9H^2a^2(\cs-w)}{k^2}\right] \right\} \sigma' +\nonumber \\
&&+ \left\{ \frac{3}{8} \frac{1+w}{\cv} + \frac{\css -w}{\css}\left[1+\frac{9H^2a^2(\cs-w)}{k^2}\right] \right\} \frac{3\sigma}{a}\nonumber \\
&& =-\frac{3}{a} \frac{\css -w}{k^2 \css}\phi_0
\eea
where we also include the term $9H^2a^2(\cs-w)/k^2$ (that we previously neglected as it is a decaying mode) 
and for simplicity we set $\phi_0 = 3H_0^2\Om\delta_0/2$ . 
The full solution then reads
\begin{widetext}
\bea
&&\sigma(a) = -\frac{\cv \phi_0 (\cs-w)}{k^2 (8 \cv
(\cs-w)+3 \cs (1+w))} \left\{8-\frac{324 \cs H_0^2 \Om (\cs-w) (1+w)}{a k^2 (4 \cv (-1+3 (\cs-w))+3
\cs (1+w))}\right. \nonumber \\
&+&\left.\frac{17496 c_{s}^4 H_0^4 \Om^2 (\cs-w)^2 (1+w)^2}{a^2 k^4 (8 \cv (-2+3 (\cs-w))+3 \cs
(1+w)) (4 \cv (-1+3 (\cs-w))+3 \cs (1+w))}\right.\nonumber \\
&-&\left.\frac{19683 c_{s}^6 H_0^6 \Om^3 (\cs-w)^3 (1+w)^3}{a^3 \cv
k^6 (-1+\cs-w) (8 \cv (-2+3 (\cs-w))+3 \cs (1+w)) (4 \cv (-1+3 (\cs-w))+3 \cs (1+w))}\right\} \nonumber \\
&+&\Big[27 \cs H_0^2 \Om (\cs-w) (1+w)+a k^2 \left[8 \cv+3 \cs (1+w)\right]\Big]^{-\frac{24
\cv (\cs-1-w)}{8 \cv+3 \cs (1+w)}}\frac{c_1}{a^3}
\label{eq:full-sigma-below}
\eea
\end{widetext}
where $c_1$ is the integration constant related to the initial conditions of the anisotropic stress $\sigma$. 
All terms inside the curly brackets except for the first one are decaying modes; moreover they all manifest a 
strong dependence on the scale $k$, becoming even less important at small scales. 
The last term in Eq.~(\ref{eq:full-sigma-below}) depends on the exponent 
\be
-\frac{24\cv (\cs-1-w)}{8 \cv+3 \cs (1+w)}\,.
\label{eq:fraction}
\ee
The former is positive if $\cv<0$ or if $\cs<1+w$; even if the last case is allowed in our calculations 
we should remember that Eq.~(\ref{eq:full-sigma-below}) was evaluated for modes below the sound horizon, 
hence for $\cs$ substantially larger than zero;  if we allow $\cs$ to be close to zero, 
then the quantity (\ref{eq:fraction}) becomes, for $w=-0.8$
\be
3(1+w)=0.6
\ee
which is a positive number; however the last term in Eq.~(\ref{eq:full-sigma-below}) is multiplied by an extra $1/a^3$ term which 
ensures us that this is still a decaying mode. 
As $\cs$ becomes larger than $1+w$, the only possibility for the last term in Eq.~(\ref{eq:full-sigma-below}) to be 
a growing mode is $\cv<0$. Again, in order to stabilise the entire solution if $\cv$ becomes negative 
one has to allow $w<-1$, {\em i.e.} phantom dark energy.

\section{Decaying modes for the density contrast}

Here we study the solutions for the dark energy density contrast using the results found in the previous section 
and verify our assumptions. We consider only modes above the sound horizon because it is the only 
case where a differential equation for the density contrast is involved. The solution is 
%\begin{widetext}
\bea
\delta(a)&=&3 \phi_0 (1 + w)^2 \left[ - \frac{1}{8 \cv k^2 w} + \frac{9 H_0^2 \Om (1 + w)}{64 c_{vis}^2 k^4 a(1+ 3  w)}\right.\nonumber \\
&+&\left.\frac{27 H_0^4 \Om^2 (1 + w)^2}{1024 a^2 c_{vis}^6 k^6 (2 + 3 w)}\right]+a^{3 w} c_2+\nonumber \\
&+&\frac{2 c_1}{\sqrt{a} H_0\sqrt{\Om}(1+ 6 w)}
\label{eq:full-deltaabove}
\eea
%\end{widetext}
where $c_1$ and $c_2$ are the constants of integration and $\phi_0 = 3H_0^2\Om\delta_0/2$. 
The second and third term in the square bracket are decaying modes and hence 
can be neglected; the second term goes like $a^{-1/2}$ and it is also a decaying mode, whereas the last term in 
Eq.~(\ref{eq:full-deltaabove}) is a growing mode if $w>0$ which is not the case of dark energy.

\section{Solutions with the effective sound horizon}\label{sec.appz-eff-cs}

In this appendix we write the solutions for the dark energy density contrast and 
velocity perturbation in terms of the effective sound speed $\ceff =\cs+8(\cs-w)/[3(1+w)]$.
\begin{itemize}
\item {\bf Modes below the sound horizon.} Eqs.~(\ref{eq:delta-sub-below}) and (\ref{eq:V-sub-below}) 
now read
\bea
\delta &=& \frac{(1+w)}{\ceff} \frac{\phi_0}{k^2}  \,,  \label{eq:delta-sub-below-ce}\\
V &=& - \frac{3(1+w)\left( \css -w \right)}{\ceff} H_0 \sqrt{\Omega_m} \frac{\phi_0}{\sqrt{a}k^2}\,.
\label{eq:v-sub-below-ce}
\eea
Here we notice that to distinguish between sound speed an the type of anisotropic stress used in 
this paper one cannot use observables which depend only on $\delta$ but one also needs to add 
observables depending on velocity perturbations.
\item {\bf Modes above the sound horizon} 
Also in this case we can write the solutions (\ref{eq:V-sub-above}) and (\ref{eq:delta-abovesh}) in 
terms of $\ceff$; however, the formers have been evaluated under the assumption $\cs\sim 0$ so that 
the effective sound speed now depends only on the viscosity term $\cv$ (and $w$)
\be
\ceff(\cs=0) = -\frac{8}{3}\frac{\cv\,w}{1+w}\,.
\ee
Then it turns out that Eqs.~(\ref{eq:V-sub-above}) and (\ref{eq:delta-abovesh}) have the exact same form 
as Eqs. (\ref{eq:delta-sub-below-ce}-\ref{eq:v-sub-below-ce}), with $\cs = 0$.
\end{itemize}

{}

\end{document}